# The non-linear health consequences of living in larger cities

Luis E C Rocha*[1,2], Anna E. Thorson[2], Renaud Lambiotte[1]

[1]Department of Mathematics and naXys, Université de Namur, Namur, Belgium
[2]Department of Public Health Sciences, Karolinska Institutet, Stockholm, Sweden
*luis.rocha@ki.se

9 June 2015

**Abstract**

Urbanization promotes economy, mobility, access and availability of resources, but on the other hand, generates higher levels of pollution, violence, crime, and mental distress. The health consequences of the agglomeration of people living close together are not fully understood. Particularly, it remains unclear how variations in the population size across cities impact the health of the population. We analyze the deviations from linearity of the scaling of several health-related quantities, such as the incidence and mortality of diseases, external causes of death, wellbeing, and health-care availability, in respect to the population size of cities in Brazil, Sweden and the USA. We find that deaths by non-communicable diseases tend to be relatively less common in larger cities, whereas the per-capita incidence of infectious diseases is relatively larger for increasing population size. Healthier life style and availability of medical support are disproportionally higher in larger cities. The results are connected with the optimization of human and physical resources, and with the non-linear effects of social networks in larger populations. An urban advantage in terms of health is not evident and using rates as indicators to compare cities with different population sizes may be insufficient.

**INTRODUCTION**

Migration from rural to urban areas reached a shift point in 2007 when the United Nations estimated that the world urban population had surpassed the rural[1]. This is a complicated phenomenon motivated not only by economic reasons such as employment opportunities and higher salaries but also a natural consequence of improved agricultural technology that demands less human resources on the field. Migration between smaller and larger cities has contributed to urbanization, and long-term projections estimate that cities will keep increasing in population size[2]. The aggregation of people in larger cities has positive and negative consequences for the society[3, 4, 5] and the so-called urban advantage[6] is not yet a consensus. While urbanization promotes innovation, develops the local economy[7, 8], optimizes resources [8], as for example the number and quality of specialized hospitals, it is also associated to higher criminal activity[9], homicides[10], carbon emissions[11, 12, 13], and mental distress[14]. These and other fundamental amenities of urban life such as efficient transportation systems or sanitation and wastewater management, all combine to the pool of factors affecting health and wellbeing[15, 16]. Efforts have been made to map the complex interconnections between all these factors aiming to develop the concept of healthy cities[15]. Complexity, however, means that the behavior of the system is not simply the sum of the behavior of its individual parts. Complex systems, like cities, may incorporate, among other characteristics, highly heterogeneous multi-level networks, fitness, feedback, and chaotic behavior[17, 18]. All combined, these characteristics may generate non-linear responses to linear inputs.

Health related quantities are characterized by a number of statistical indicators to summarize features of a given population aiming to rank and compare individual cities. These indicators are widely used by policy makers to identify areas that need further attention or to quantify and monitor achievements of previous efforts. Well-known indicators include life expectancy, death and prevalence rates, income, coverage, and so on. Many times, a city-level indicator such as a rate is defined by dividing a quantity of interest by the population size of the respective city, sometimes presented as the number of cases per number of individuals, e.g. 100,000 or 1 (per-capita), within a given time interval. The procedure to define a rate implicitly assumes a linear relation between the dependent and the independent variables[9]. Technically, this linear assumption allows an unbiased per-capita analysis only if the quantity under observation is size-independent. On the other hand, the non-linearity, represented by the parameter $\alpha$ in equation (1), means that cities with larger populations $P$ observe a relative per-capita increase ($\alpha > 1$) or decrease ($\alpha < 1$) of the variable $Q$. The parameter $\beta$ is a constant and does not affect the linear scaling.

$$Q = \beta P^{\alpha} \qquad (1)$$

The scaling parameter $\alpha$ thus indicates the relative advantage of living in larger cities by capturing the deviations from the linear scaling. The use of equation (1) is motivated by the study of power-law scaling between characteristics of living creatures and their body size. Allometry dates back to 1936[19] but more recently it has been adapted to study the relation between characteristics of cities and their population through a metaphor between cities and living creatures[8]. While some theories have been proposed to justify power-law relations in biological systems[20], a robust theory of scaling for cities is still missing. The estimation of these scaling relations contribute to understand the macroscopic impact of the population size of cities in health[21], particularly to rank cities across a country. It also reflects the consequences of the increasing complexity of daily interactions between people and the available resources observed in cities with larger populations, that eventually affect health variables[22].

In this paper, we analyze the scaling laws of several health related variables in Brazil, Sweden, and the USA. These countries have different income levels, wealth distribution, cultural background, and health care and governmental policies. The lack of standardized data for multiple countries restricts the generalizability of the results and thus caution should guide causality associations between the scaling of the different variables particularly because some results are country-specific. Nevertheless, whenever possible we make a comparative analysis between the different contexts. We particularly look at the prevalence of several infectious and non-communicable diseases, causes of death, children related health and social variables, wellbeing, and availability and access to health care facilities. We emphasize however that results are not necessarily universal but country-specific.

**MATERIALS AND METHODS**
**Datasets**

For each country, we collect city-level data on the variable of interest and on the population size. To avoid misinterpretation of the concept of city, we simply use the finest resolution of urban area as available in the respective datasets. This translates to nearly 5,550 places in Brazil, 210 in Sweden and 980 in the USA. Note that the exact number of cities (i.e. data points) depends on the particular study variable since cities with no reported cases are removed for the statistics. There is also no standardization across countries and we use the definition of the variables as they are described on each dataset. We consider the place of residence and not the local of occurrence for the statistics. The Brazilian dataset is obtained from "Data SUS", that is the department of informatics of the Brazilian publicly funded health care system (www2.datasus.gov.br). The Swedish dataset comes from "Folkhälsomyndigheten", the public health agency of Sweden (www.folkhalsomyndigheten.se). The datasets related to the United States of America (USA) are obtained from "The County Health Rankings & Roadmaps program", a consortium between the Robert Wood Johnson Foundation and the University of Wisconsin Population Health Institute (www.countyhealthrankings.org).

**Analysis**

We initially construct two vectors $Q$ and $P$, which entries $q_i$ and $p_i$ contain, respectively, the variable of interest and the population size of the city $i$. To estimate the coefficients $\alpha$ and $\beta$, we first take the logarithm on both sides of equation (1).

$$log(Q) = log(\beta) + \alpha\, log(P) \quad (2)$$

to linearize the original non-linear formula such that $A=log(\beta)$ and $B=\alpha$ in $y=A + Bx$. Typically the range of values of the population sizes spans several orders of magnitude. The logarithm operator reduces the difference in magnitude of these values and allows the estimation of the parameters using a simple regression analysis on equation (2). Alternatively, one may apply more advanced methods, e.g. weighted least squares, directly on equation (1)[9]. We fit the data-points in the intervals [50,000–5,000,000] (for Brazil and the USA) and [10,000–500,000] (for Sweden) of population size to remove the small cities, that typically cause a large dispersion for low values, and to remove very large cities, that are statistically underrepresented. If no cases are observed in a city for a certain variable, the city is not used for the statistics. This procedure is standard in the literature[8, 23]. The exponent $\alpha$ and $\beta$, the respective confidence intervals, and the adjusted correlations coefficients $R^2$ are estimated using the statistical software R (www.r-project.org).

**Super- vs. sub-linear scaling**

The non-linear dependence between variables is not as intuitive as the linear relation. Mathematically non-linearity is characterized by either the so-called super-linear ($\alpha > 1$) or the sub-linear ($\alpha < 1$) scaling exponent that measures the deviation from the linear behavior ($\alpha = 1$) in equation (1). Note that the parameter $\beta$, that is a constant, measures the intensity of the relation between the variables and is not relevant to determine the non-linearity. Using equation (1), one can show that the relative increase in a given outcome $Q$ for an increase $P_2 = \delta P_1$ in the population is $(Q_2-Q_1)/Q_1 = \delta^\alpha - 1$, where $\delta$ is a constant number and $Q_1$ and $Q_2$ correspond to the value of the variable $Q$ (e.g. number of cases, counts) associated respectively to the population size $P_1$ and $P_2$. Therefore, if the dependence between the two variables is linear, $\alpha = 1$ and $Q_2 = \delta Q_1$. On the other hand, in case of non-linear scaling, i.e. $\alpha \neq 1$, the relative increase becomes non-trivial. To understand quantitatively this difference, Figure 1a shows the relative quantity $(Q_2-Q_1)/Q_1$ for a given change $\delta$ in the population size, for different values of the exponent $\alpha$. To facilitate the analysis, we arbitrarily define ranges for the deviations to the linear behavior based on typical results reported in the literature: $0.95 \leq \alpha \leq 1.05$ (weak deviation), $0.85 \leq \alpha < 0.95$ and $1.05 < \alpha \leq 1.15$ (medium), and $\alpha < 0.85$ and $\alpha > 1.15$ (strong). As an example, if $\alpha = 1.1$ and the population increases by 100% (i.e. $\delta = 2$), the outcome increases by 114.35%. On the other hand, for $\alpha = 0.9$, the same 100% increase in the population causes an increase of only 86.61% in the outcome $Q$. In this particular case, the nearly 15% difference, accounts for a mismatch (in respect to the linear assumption) of 1,500 in 10,000 counts. This example also illustrates the non-symmetry of the exponent around its linear value, i.e. the relative increase given by $α = 1.1$ is not equivalent to the relative decrease given by $α = 0.9$.

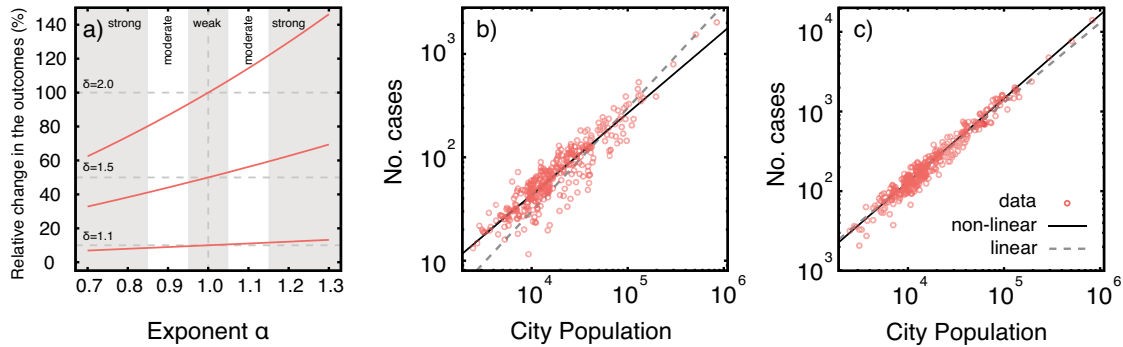

Figure 1. a) Relative change in the outcome $Q$, $(Q_2-Q_1)/Q_1$ (y-axis), for different values of the exponent $α$ (x-axis). For $α = 1$, we recover the usual linear relation between population size and the outcome variable. The relation between the population size (x-axis) and b) the number of deaths by heart attack or the c) number of abortions (y-axis). Both axes are in log-scale.

**RESULTS**

To illustrate a typical scale relation between a variable and the population size of a city, we show a scatter plot of the population size of cities and the number of deaths by heart attack (Figure 1b) and the same relation for the number of abortions (Figure 1c) in the context of Sweden. The deviation to the linear trend is not visually evident but the number of heart attacks scales strongly sub-linearly ($α = 0.80$ with 95% C.I. [0.74,0.86]) whereas the number of abortions scales weakly super-linearly ($α = 1.05$ with 95% C.I. [1.01,1.09]) with the population size.

For infectious diseases, the exponent $α$ is typically larger than one (Table I), being strongly super-linear for sexually transmitted infections[24], meningitis and the 2009 pandemic influenza in Brazil. Infectious diseases spread mainly through contacts between an infected host and a healthy person[25]. At the city level, the high connectivity of the human contact networks, which scales super-linearly with city size[26], facilitates the spread of infections and may be the responsible for these strong deviations. The fact that the 2010 pandemic influenza scales linearly with city population possibly indicates an effective response after the epidemic outbreak in the previous year. It has been reported that in the influenza pandemic of 1918, the number of deaths scale sub-linearly with the population size of US cities. On the other hand, the same authors report that pneumonia deaths follow a linear relation[27]. The sub-Linear scaling of leprosy cases indicates that smaller cities are relatively more affected than larger cities. This may be related to the fact that leprosy, although with known cure, is a silent, sometimes difficult to diagnose disease[28], and therefore, disproportionally affect areas with insufficient medical resources. Dengue, on the other hand, spreads through an infected mosquito that is typically prevalent in areas lacking proper sanitation and mosquito control. Although urbanized areas (e.g. slums) are associated to increased risk of infection[29], our analysis suggests that in Brazil, the risk increases linearly with the population size. We have also identified that the number of deaths by external causes and by non-communicable diseases typically deviates from the linear behavior (Table I). The results indicate that larger cities have relatively less deaths by external causes such as car accidents, injury or suicides[23]. Deaths by non-external causes (non-communicable diseases) depend on the nature of the disease and on the country. These results possibly reflect not only the local behavior (e.g. healthier life style) but also an equally distributed and accessible health care system. For example, diabetes deaths and heart attacks are both strongly sub-linear in Sweden but are respectively weakly sub-linear and weakly super-linear in Brazil in 2012. In comparison to previous years, however, both indicators have decreased in Brazil. Lung cancer and chronic respiratory insufficiency are also sub-linear in Sweden but consistently super-linear in Brazil. The population size of cities may also affect children and families. We observe a relative decrease in mortality for both infants and children in larger cities, but child poverty vary from weakly sub-linear in the USA to weakly super-linear in Sweden (Table I). According to Swedish data, there is a small tendency to find medium size families in larger cities, where pregnant women are also more likely to perform an abortion.

| Indicator | Country | Period | α | 95% C.I. | β | 95% C.I. | Adj. $R^2$ |
|---|---|---|---|---|---|---|---|
| **Incidence of Infectious Diseases** | | | | | | | |
| HIV | BRA | 1990 | 1.20 | [1.07,1.34] | 0 | [-4.41,4.41] | 0.52 |
| HIV | BRA | 2012 | 1.38 | [1.31,1.45] | 0 | [-3.00,3.00] | 0.71 |
| HIV | USA | 2000s | 1.44 | [1.39,1.50] | 0 | [-2.69,2.69] | 0.76 |
| Chlamydia | USA | 2011 | 1.14 | [1.11,1.18] | 0 | [-2.44,2.44] | 0.80 |
| Pandemic Influenza | BRA | 2009 | 1.20 | [1.04,1.37] | 0 | [-5.28,5.28] | 0.29 |
| Pandemic Influenza | BRA | 2010 | 1.00 | [0.89,1.11] | 0 | [-3.88,3.88] | 0.41 |
| Meningitis | BRA | 2001 | 1.25 | [1.16,1.34] | 0 | [-3.44,3.44] | 0.63 |
| Meningitis | BRA | 2012 | 1.39 | [1.26,1.52] | 0 | [-4.28,4.28] | 0.51 |
| Hepatite A | BRA | 2007 | 1.08 | [0.93,1.23] | 0 | [-4.81,4.81] | 0.36 |
| Hepatite A | BRA | 2012 | 1.06 | [0.93,1.19] | 0 | [-4.41,4.41] | 0.37 |
| Hepatite B | BRA | 2007 | 0.98 | [0.86,1.09] | 0 | [-3.93,3.93] | 0.38 |
| Hepatite B | BRA | 2012 | 1.06 | [0.95,1.17] | 0 | [-3.76,3.76] | 0.43 |
| Hepatite C | BRA | 2007 | 1.08 | [0.93,1.23] | 0 | [-4.81,4.81] | 0.36 |
| Hepatite C | BRA | 2012 | 1.06 | [0.93,1.19] | 0 | [-4.41,4.41] | 0.37 |
| Dengue | BRA | 2001 | 0.98 | [0.70,1.25] | 0 | [-9.92,9.92] | 0.09 |
| Dengue | BRA | 2012 | 1.03 | [0.82,1.24] | 0 | [-6.95,6.95] | 0.14 |
| Leprosy | BRA | 2001 | 0.89 | [0.75,1.04] | 0 | [-4.60,4.60] | 0.23 |
| Leprosy | BRA | 2012 | 0.77 | [0.64,0.90] | 0 | [-4.35,4.35] | 0.18 |
| **External Causes of Death** | | | | | | | |
| Car Accidents | BRA | 1981 | 1.13 | [1.03,1.22] | 0 | [-3.47,3.47] | 0.60 |
| Car Accidents | BRA | 2012 | 0.89 | [0.84,0.94] | 0 | [-2.59,2.59] | 0.70 |
| Alcohol Imp Driving | USA | 2008-12 | 0.71 | [0.67,0.74] | 0 | [-2.45,2.46] | 0.59 |
| Injury | USA | 2006-10 | 0.89 | [0.87,0.91] | 0.01 | [-2.19,2.21] | 0.90 |
| Suicides | BRA | 1981 | 0.84 | [0.74,0.95] | 0 | [-3.68,3.68] | 0.46 |
| Suicides | BRA | 1995 | 0.92 | [0.84,1.01] | 0 | [-3.32,3.33] | 0.50 |
| Suicides 65+ | SWE | 2008-12 | 0.86 | [0.79,1.00] | 0 | [-3.37,3.37] | 0.59 |
| Suicides 65- | SWE | 2008-12 | 0.95 | [0.87,1.03] | 0 | [-2.94,2.95] | 0.73 |
| **Non-external Causes of Death** | | | | | | | |
| Diabetes | BRA | 1996 | 1.22 | [1.13,1.31] | 0 | [-3.30,3.30] | 0.61 |
| Diabetes | BRA | 2012 | 0.97 | [0.92,1.02] | 0 | [-2.65,2.65] | 0.70 |
| Diabetes | SWE | 2008-12 | 0.77 | [0.68,0.86] | 0 | [-3.18,3.19] | 0.56 |
| Heart Attack | BRA | 1981 | 1.25 | [1.12,1.38] | 0 | [-4.29,4.29] | 0.58 |
| Heart Attack | BRA | 2012 | 1.04 | [0.98,1.09] | 0 | [-2.71,2.71] | 0.70 |
| Heart Attack | SWE | 2008-12 | 0.80 | [0.74,0.86] | 0.03 | [-2.66,2.71] | 0.76 |
| Cerebrovasc. accident | BRA | 1996 | 1.16 | [1.08,1.25] | 0 | [-3.24,3.24] | 0.60 |
| Cerebrovasc. accident | BRA | 2012 | 1.00 | [0.97,1.04] | 0 | [-2.44,2.44] | 0.83 |
| Lung cancer | BRA | 1981 | 1.14 | [1.02,1.25] | 0 | [-3.86,3.86] | 0.57 |
| Lung cancer | BRA | 2012 | 1.16 | [1.09,1.22] | 0 | [-2.95,2.96] | 0.65 |
| Lung cancer | SWE | 2008-12 | 0.94 | [0.90,0.99] | 0 | [-2.49,2.49] | 0.89 |
| Chronic resp insuf. | BRA | 1981 | 1.14 | [1.02,1.25] | 0 | [-3.90,3.90] | 0.53 |
| Chronic resp insuf. | BRA | 2012 | 1.07 | [1.00,1.14] | 0 | [-2.99,2.99] | 0.59 |
| Chronic resp insuf. | SWE | 2008-12 | 0.90 | [0.85,0.95] | 0 | [-2.56,2.56] | 0.86 |
| Colon cancer | BRA | 1981 | 0.77 | [0.66,0.88] | 0 | [-3.78,3.78] | 0.46 |
| Colon cancer | BRA | 2012 | 1.00 | [0.94,1.07] | 0 | [-2.88,2.88] | 0.64 |
| **Infant and children related quantities** | | | | | | | |
| Infant mortality (0-1year) | BRA | 1981 | 1.10 | [0.98,1.22] | 0 | [-3.99,3.99] | 0.48 |
| Infant mortality (0-1year) | BRA | 2012 | 0.98 | [0.94,1.03] | 0 | [-2.53,2.53] | 0.77 |
| Infant mortality (0-1year) | USA | 2000s | 0.88 | [0.87,0.90] | 0.04 | [-2.15,2.24] | 0.90 |
| Child mortality (1-9years) | BRA | 1981 | 1.02 | [0.91,1.12] | 0 | [-3.65,3.65] | 0.51 |
| Child mortality (1-9years) | BRA | 2012 | 0.94 | [0.88,1.00] | 0 | [-2.84,2.84] | 0.60 |
| Child mortality (1-9years) | USA | 2000s | 0.98 | [0.95,1.00] | 0 | [-2.26,2.26] | 0.88 |
| Child poverty (18-) | USA | 2000s | 0.97 | [0.94,1.00] | 0.07 | [-2.28,2.42] | 0.81 |
| Child poverty | SWE | 2012 | 1.03 | [1.01,1.05] | 0.15 | [-2.05,2.34] | 0.98 |
| Abortion | SWE | 2007-11 | 1.05 | [1.01,1.09] | 0 | [-2.38,2.38] | 0.94 |
| Family 0 children | SWE | 2013 | 0.99 | [0.97,1.02] | 0.42 | [-1.79,2.62] | 0.97 |
| Family 1 or 2 children | SWE | 2013 | 1.05 | [1.03,1.07] | 0.06 | [-2.12,2.23] | 0.98 |
| Family 3+ children | SWE | 2013 | 1.00 | [0.97,1.03] | 0.02 | [-2.27,2.30] | 0.96 |

Table I. Scaling exponent α and constant β for the prevalence of infectious diseases, number of deaths by external and non-external (non-communicable diseases) causes, and children related health and social variables.

| Indicator | Country | Period | α | 95% C.I. | β | 95% C.I. | Adj. $R^2$ |
|---|---|---|---|---|---|---|---|
| **Personal behavior** | | | | | | | |
| Drug Poisoning | USA | 2000s | 0.99 | [0.96,1.02] | 0 | [-2.38,2.38] | 0.80 |
| Smokers | USA | 2006-12 | 0.89 | [0.87,0.90] | 72.20 | [70.02,74.38] | 0.91 |
| Physically Inactive** | USA | 2010 | 0.92 | [0.90,0.93] | 67.10 | [64.96,69.23] | 0.95 |
| Access Exercise Opportunity | USA | 2010-12 | 1.20 | [1.18,1.22] | 6.04 | [3.84,8.24] | 0.94 |
| Excessive Drinking | USA | 2006-12 | 1.06 | [1.04,1.08] | 7.50 | [5.28,9.72] | 0.92 |
| Alcohol Sales | SWE | 2012 | 0.98 | [0.89,1.07] | 5.26 | [2.14,8.37] | 0.69 |
| BMI 18.5- | SWE | 2010-13 | 1.12 | [1.04,1.20] | 0.51 | [-2.50,3.52] | 0.85 |
| BMI 18.5-24.9 | SWE | 2010-13 | 1.07 | [1.05,1.09] | 22.19 | [20.01,24.36] | 0.99 |
| BMI 25-29.9 | SWE | 2010-13 | 0.95 | [0.93,0.96] | 63.34 | [61.21,65.47] | 0.99 |
| BMI 30+ | SWE | 2010-13 | 0.89 | [0.85,0.92] | 47.49 | [45.10,49.87] | 0.94 |
| BMI 30+ | USA | 2010 | 0.93 | [0.92,0.94] | 63.76 | [61.67,65.85] | 0.97 |
| Lim Healthy Food | USA | 2000s | 0.87 | [0.82,0.92] | 0.27 | [-2.39,2.93] | 0.54 |
| Diabetic Medicare Enrollees | USA | 2011 | 0.93 | [0.92,0.95] | 21.81 | [19.67,23.95] | 0.95 |
| Household Severe Problems* | USA | 2006-10 | 1.10 | [1.09,1.12] | 0.02 | [-2.15,2.18] | 0.95 |
| Lack Emo Support | SWE | 2010-13 | 1.00 | [0.97,1.03] | 11.32 | [8.97,13.68] | 0.96 |
| Lack Pract Support | SWE | 2010-13 | 1.07 | [1.01,1.12] | 2.45 | [0.25,5.14] | 0.90 |
| Red Psycho. Wellbeing | SWE | 2010-13 | 1.10 | [1.08,1.13] | 5.20 | [2.91,7.48] | 0.98 |
| Drive Alone Work | USA | 2008-12 | 0.98 | [0.97,0.99] | 0.43 | [-1.67,2.53] | 0.97 |
| **External Causes of Injuries** | | | | | | | |
| Motor Accidents | USA | 2000s | 0.75 | [0.72,0.77] | 0.02 | [-2.26,2.30] | 0.78 |
| Alcohol Related Crimes | SWE | 2013 | 0.93 | [0.86,1.01] | 0 | [-2.87,2.87] | 0.75 |
| Violent Crimes | USA | 2009-11 | 1.22 | [1.18,1.26] | 0 | [-2.58,2.58] | 0.75 |
| Rapes | BRA | 2009 | 0.92 | [0.80,1.03] | 0 | [-4.03,4.03] | 0.45 |
| Rapes | BRA | 2012 | 1.29 | [1.19,1.38] | 0 | [-3.56,3.56] | 0.55 |
| Rapes | SWE | 2013 | 1.14 | [1.03,1.26] | 0 | [-3.55,3.55] | 0.66 |
| Dom Phys Violence | BRA | 2009 | 0.92 | [0.80,1.03] | 0 | [-4.03,4.03] | 0.45 |
| Dom Phys Violence | BRA | 2012 | 1.28 | [1.13,1.43] | 0 | [-4.82,4.82] | 0.32 |
| Dom Moral Violence | BRA | 2009 | 1.03 | [0.87,1.19] | 0 | [-5.12,5.12] | 0.35 |
| Dom Moral Violence | BRA | 2012 | 1.18 | [1.04,1.32] | 0 | [-4.53,4.53] | 0.34 |
| **Health-care Availability and Access** | | | | | | | |
| Physicians | BRA | Dec 2012 | 1.45 | [1.38,1.53] | 0 | [-3.03,3.03] | 0.72 |
| Primary Care Physicians | BRA | Dec 2012 | 1.28 | [1.20,1.35] | 0 | [-3.06,3.06] | 0.65 |
| Primary Care Physicians | USA | 2011 | 1.18 | [1.15,1.21] | 0 | [-2.34,2.34] | 0.87 |
| Dentists | BRA | Dec 2012 | 1.19 | [1.12,1.26] | 0 | [-2.96,2.96] | 0.65 |
| Dentists | USA | 2011 | 1.22 | [1.19,1.24] | 0 | [-2.31,2.31] | 0.89 |
| Mental Health Providers | USA | 2013 | 1.31 | [1.26,1.36] | 0 | [-2.69,2.69] | 0.72 |
| Mental Health Providers | BRA | Dec 2012 | 1.29 | [1.20,1.37] | 0 | [-3.20,3.20] | 0.61 |
| Nutritionists | BRA | Dec 2012 | 1.21 | [1.14,1.29] | 0 | [-3.01,3.01] | 0.65 |
| Primary Care Clinic | BRA | Dec 2012 | 1.10 | [1.03,1.17] | 0 | [-2.95,2.95] | 0.62 |
| Private Insurance | BRA | Dec 2000 | 1.64 | [1.51,1.76] | 0 | [-4.07,4.07] | 0.57 |
| Private Insurance | BRA | Dec 2012 | 1.69 | [1.59,1.79] | 0 | [-3.63,3.63] | 0.63 |
| Insured (65+ years) | USA | 2011 | 1.00 | [1.00,1.00] | 0.87 | [-1.14,2.87] | 0.99 |
| Medicare Enrollees | USA | 2000s | 0.83 | [0.80,0.85] | 0.06 | [-2.19,2.31] | 0.84 |
| Not visited a doctor | USA | 2000s | 0.96 | [0.93,0.98] | 22.12 | [19.85,24.39] | 0.87 |
| Had contact with health services | SWE | 2010-13 | 1.00 | [0.98,1.01] | 55.66 | [53.55,57.77] | 0.99 |
| Had not visited dentist previous 2 years | SWE | 2010-13 | 1.01 | [0.96,1.06] | 10.02 | [7.46,12.59] | 0.92 |
| Gross GDP | BRA | 2000 | 1.27 | [1.19,1.35] | 0.21 | [-2.92,3.34] | 0.66 |
| Gross GDP | BRA | 2010 | 1.30 | [1.23,1.37] | 0.43 | [-2.54,3.41] | 0.69 |

Table II. Scaling exponent α and constant β for variables related to wellbeing, behavior, availability and access of health-care facilities, and external causes of injuries. *households with at least 1 out of 4 housing problems e.g. overcrowding, high housing costs, lack of kitchen or plumbing facilities, **no leisure time physical activity.

Our results suggest that people have a tendency to healthier life style in larger cities in the USA, for example we observe substantially less smoking and physical inactivity, and more access to healthy food (Table II). Similarly, obesity is also proportionally less common in larger cities in both Sweden and the USA. In contrast, limited data[30] suggest that smoking is disproportionally higher in larger capital cities of Brazil ($\alpha$ =1.05 with 95% C.I. [0.95,1.15]; $\beta$=6.68 with 95% C.I. [2.68,10.68]). Such activities have a strong social component. Recent studies suggest that the social networks affect obesity[31, 32] and smoking habits[33], not only by peer fitness (or peer-attraction) but also by influencing one's close contacts, as for example close relatives. On the other hand, citizens of larger cities are more exposed to violent crimes and have reported a disproportionally higher number of rapes (Brazil and Sweden) and household problems, including domestic violence (Brazil). In some contexts, city life also causes distress on the population reducing the psychological wellbeing (Sweden) and increasing excessive drinking (USA).

Larger cities are known to disproportionally generate knowledge and capital, to optimize resources[8], and to boost productivity due to higher social interactions[34, 35]. The optimization of resources also implies that cities are expected to concentrate high-quality specialized health care centers and professionals. For various diseases however generalist clinics and physicians are sufficient to provide appropriate primary care. Health care providers (primary care, dentists, mental health, and nutritionists) are more abundant per-capita in larger cities both in the USA and in Brazil (Table II). If we compare the scaling for the case of all (specialized and primary care) physicians and only primary care physicians in Brazil, we see that larger cities concentrate a disproportional per-capita number of specialists. The number of privately insured individuals is also strongly super-linear in Brazil (health care is public and completely free in Brazil) whereas in the USA the elderly is, on average, equally insured independently of the city size. Medicare enrollees (social insurance program for elderly and young people with disabilities in the USA), on the other hand, are more concentrated in smaller cities. In Sweden, visits to physicians and dentists are equally distributed across the cities. In general, we observe that access to different forms of health care services is relatively more abundant in larger cities. These facts may be connected to the relative lower burden of non-communicable diseases, suicides, and child and infant mortality observed in larger cities (Table II) but stronger causal conclusions depend on more carefully analysis of country-specific data.

The exponent $\alpha$ = 1 indicates a relative equilibrium between the respective variable and the population size across cities within a country. For example, doubling the population results on two times more problems or benefits. This equilibrium is not necessarily optimal because it may indicate that physical resources may be under-exploited, as for example, facilities for physical activity or number of beds in hospitals. While there is no clear stable or optimal point for this exponent, deviations from the linearity suggest that resources should not be proportionally allocated if one wishes to promote a similar quality of life to the entire country.

Efforts to promote equality may thus be reflected in the evolution of the scaling exponents, which are not expected to be constant in time. Considering the Brazilian context, Figure 2 shows that HIV affected smaller cities more severely in the early stage of the epidemics in the 1980s. As the epidemics pervaded the population, larger cities have disproportionally suffered. In other words, HIV more easily spread in larger cities. Infant mortality, diabetes and heart attacks have all approached linear scaling in recent years. For diabetes and heart attacks, it is difficult to infer if the improvement is a result of better health in larger cities or better diagnosis in smaller cities. Deaths by lung cancer have maintained super-linear exponents over the 30-year period. On the other hand, deaths by colon cancer have affected relatively more people in larger cities in recent years than in the 1980s. Convergence to linear or sub-linear exponents in these cases is a positive sign of an increasing urban advantage.

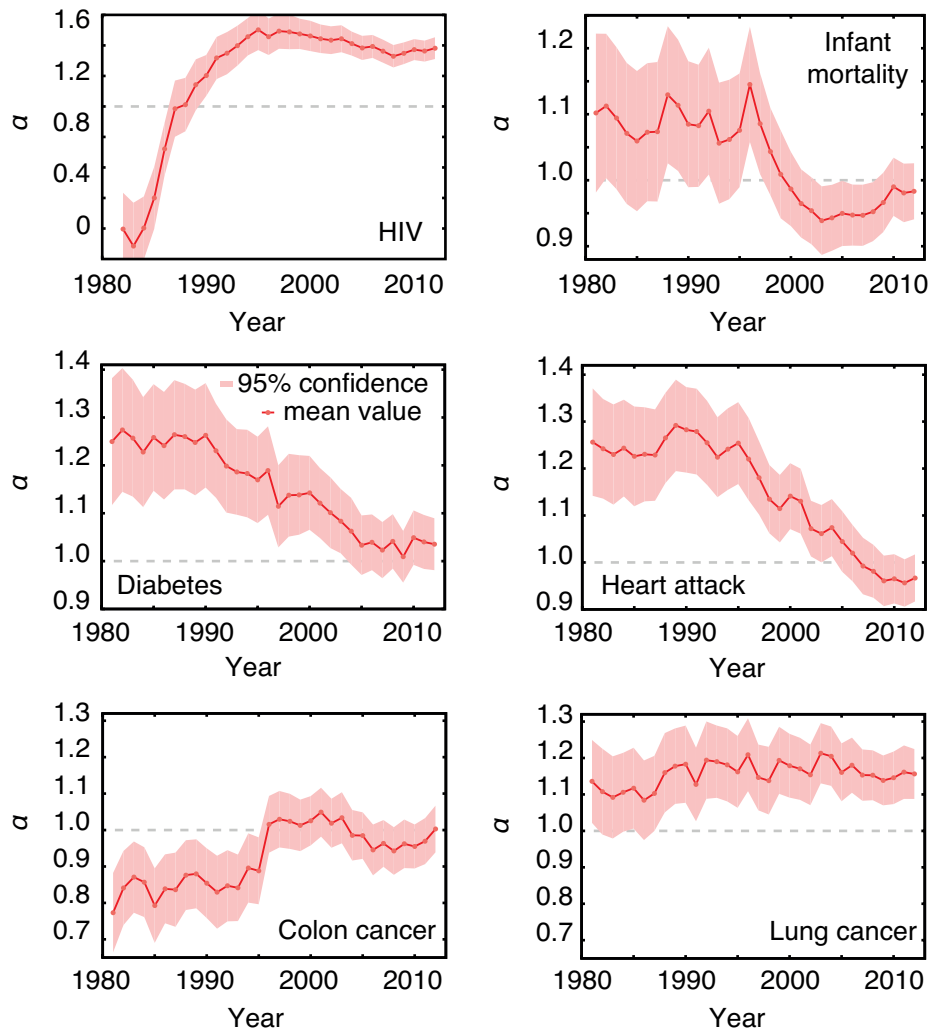

Figure 2: The panels show the variation of the scaling exponent α (together with the 95% C.I. – shade area) for selected variables in the period from 1981 to 2012 in Brazil. Dotted horizontal lines correspond to the reference of linear scaling α = 1.

We have been looking at the scaling properties of different cities at a given year. These analyses provide comparative information at the national level but little say regarding the development of a single city. The health variables discussed above are not expected to change equally across cities when the cities increase in size. Each city in fact goes through a different dynamics over time (e.g. investments, education) that affects health in one-way or another even in cities with similar size (a phenomenon known as the Glasgow effect [36]). If we perform the previous analysis for single cities in Brazil at different years, we observe that the scaling properties vary substantially for some health variables in selected cities of different sizes (Table III).

HIV is a particular case because the epidemics started in the early 1980s, experiencing a substantial growth in the early years in some cities, followed by a slower but still high incidence after the peak of the epidemics in the early 1990s. Infant mortality on the other hand had a negative strong variation in the exponent $\alpha$, a positive sign reflecting a strong decrease in the number of deaths over the years. Only for the city of Belo Horizonte that we observe a sub-linear exponent for diabetes whereas the number of heart attacks varies for each city. The number of cases of colon and lung cancer is in general strongly super-linear with the exception of colon cancer in Campinas ($\alpha = -1.77$) and Belo Horizonte ($\alpha = 0.32$).

| | Large-size cities about 2,500,000 inhabitants | | | |
|---|---|---|---|---|
| | Fortaleza | | Belo Horizonte | |
| | $\alpha$ | 95% C.I. | $\alpha$ | 95% C.I. |
| HIV | 20.1 | [17.7,22.4]♭ | 3.25 | [1.82,4.69]♮ |
| | 4.18 | [3.48,4.88]✤ | 7.89 | [5.72,10.0]• |
| Infant Mortality | -3.06 | [-3.42,-2.69] | -7.84 | [-8.41,-7.27] |
| Diabetes | 2.73 | [2.42,3.05] | 0.89 | [0.21,1.57] |
| Heart Attack | 0.42 | [0.18,0.65] | 0.55 | [0.17,0.93] |
| Colon Cancer | 2.83 | [2.33,3.33] | 0.32 | [-0.40,1.05] |
| Lung Cancer | 2.73 | [2.49,2.97] | 2.65 | [2.29,3.00] |
| | Medium-size cities about 500,000 inhabitants | | | |
| | Campinas | | Sao Luis | |
| | $\alpha$ | 95% C.I. | $\alpha$ | 95% C.I. |
| HIV | 30.7 | [24.5,36.8]♭ | 7.60 | [6.85,8.35] |
| | 0.94 | [0.18,1.70]✤ | | |
| Infant Mortality | -3.10 | [-3.39,-2.81] | -1.68 | [-1.90,-1.46] |
| Diabetes | 1.48 | [1.11,1.85] | 2.88 | [2.62,3.15] |
| Heart Attack | 1.80 | [1.46,2.15] | 0.45 | [-0.02,0.92] |
| Colon Cancer | -1.77 | [-2.69,-0.86] | 3.68 | [2.54,4.83] |
| Lung Cancer | 1.99 | [1.62,2.36] | 1.91 | [1.67,2.16] |
| | Small-size cities about 100,000 inhabitants | | | |
| | Londrina | | Cuiaba | |
| | $\alpha$ | 95% C.I. | $\alpha$ | 95% C.I. |
| HIV | 27.3 | [14.9,39.7]♮ | 5.50 | [4.78,6.27] |
| | 1.49 | [0.68,2.29]• | | |
| Infant Mortality | -2.96 | [3.29,-2.62] | -0.52 | [-1.08,0.03] |
| Diabetes | 3.63 | [2.75,4.52] | 4.17 | [3.70,4.64] |
| Heart Attack | 0.17 | [-0.34,0.68] | 1.62 | [1.26,1.99] |
| Colon Cancer | 2.19 | [1.28,3.11] | 2.19 | [1.28,3.11] |
| Lung Cancer | 3.12 | [2.68,3.57] | 2.25 | [1.87,2.63] |

Table III. Scaling exponent $\alpha$ for single cities, estimated using the population size and the number of cases for each city at different years. The HIV data was split in two time periods, therefore the symbols ♭, v, ♮, w mean respectively more than 200 cases, less than 200 cases, more than 50 cases and less than 50 cases of HIV.

## DISCUSSIONS

By analyzing the scaling relation between the population size of cities and various health related variables, we find that in general those conditions directly involving interaction between people scales super-linearly with the population size. This is particularly the case of various infectious diseases, with an exception of those linked to resource-poor environments such as dengue and leprosy. This means that larger cities have a relatively higher incidence of infectious diseases that translates into a relative disadvantage in respect to smaller cities. These results suggest that infectious diseases spread relatively faster within denser agglomeration of people, possibly a result of the non-linear contact patterns between individuals (people make a higher number of per-capita contacts in larger cities[26]), and the human mobility patterns[37]. Similarly, the number of violent crimes, rapes, domestic violence and household problems increase super-linearly with population. On the other hand, the number of deaths by suicides, heart attacks, or cases of diabetes is either sub-linear or close to linear depending on the country studied. This means that the number of deaths by these factors is relatively smaller in larger cities, that is, large city dwellers are relatively less likely to commit suicide, die by a heart attack or develop diabetes than those living in smaller cities. This could be linked to the relative disproportion of elderly people living in smaller cities[38] but strong conclusions require further studies. These results also relate to our findings that obesity (USA and Sweden), smoking (USA), physical inactivity (USA), and limited access to health food (USA) all scale sub-linearly with population size considering the available data, indicating that relatively speaking dwellers of larger cities have on average a healthier life style, if only those variables are considered, possibly benefiting of the social networks[39]. Preliminary data in the context of Brazil however indicate that smoking increases super-linearly with population size, suggesting that healthy life style may not be a universal characteristic of larger cities. Furthermore, previous studies found that the incidence of psychosis and depression increases with the population density[40, 41] suggesting that mental health is more sensitive to urban living. Lung cancer and chronic respiratory insufficiency are both relatively more likely in larger cities of Brazil whereas they are relatively less likely in larger cities of Sweden. Lung cancer is linked to smoking that is weakly super-linear in Brazil. Pollution, relatively higher in larger cities, also plays a role on these diseases and may, at least partially, explain these findings[42]. More robust data for the various countries are necessary before making stronger conclusions, at the macroscopic level, between pollution, smoking, lung cancer, and population size of cities.

Previous empirical studies show that the density of ambulatory hospitals in the USA and primary care clinics in South Korea scales super-linearly with the population density within a certain area. South Korea hospitals on the other hand scale slightly sub-linearly[43]. These results indicate that primary care facilities have a high availability in densely populated areas. Similarly, we find that in both Brazil and the USA, the availability of specialist and primary care physicians, dentists, mental health providers and nutritionists all scale strongly super-linearly with the population size. Primary care clinics and people with private insurance also increase super-linearly in Brazil. Altogether, these results show that the availability of health care per-capita is significantly higher in larger cities in those countries. Although larger cities are attraction poles of medical services for dwellers of smaller cities, the easier and increasing access to medical facilities in larger cities may explain why the number of deaths by non-external causes is continuously getting less super-linear in Brazil.

The most important limitation of this methodology is that we are unable to make causal relations between population size and health outcomes. We cannot directly use the non-linearity of one quantity to explain the scaling of another. Moreover, while several scaling exponents are similar to different countries, sometimes they give contrasting results suggesting that intrinsic characteristics of each country, such as public policies, income, cultural aspects, or even the country own geography, may be defining the health outcomes and limiting the universality of the results. Brazil, Sweden and the USA are relatively different countries yet they share similarities. If data from other countries, particularly those in the low-income bracket, become available, a theory on non-linear scaling of health variables could be further developed. Non-standardized and missing data also restrict a detailed comparison across countries. Recent studies have questioned the definition of city boundaries and thus the non-linear scaling with population size for some urban indicators[13, 44, 45]. Altogether, these results however help to understand the effects of the increasing complexity of larger cities in public health. For some diseases, higher per-capita efforts should be given to larger cities whereas for other diseases, efforts should target smaller cities. Another important lesson is that ranks based on simple rates miss the effects of population size and thus provide an incomplete picture of the state of health of individual cities across a country. One should therefore also look at the deviations from linearity in order to more accurately estimate trends and rank cities.